# Composition Dependent Electrochemical Properties of Earth-Abundant Ternary Nitride Anodes

M. Brooks Tellekamp[1][1], Anna Osella[1], Karen N. Heinselman[1], Adele C. Tamboli[1], Chunmei Ban[1,2]

[1] National Renewable Energy Laboratory, 15301 Denver West Parkway, Golden, CO 80401

[2] Paul M Rady Department of Mechanical Engineering, University of Colorado Boulder, 1111 Engineering Dr, UCB 427, Boulder, CO, 80209

[1] brooks.tellekamp@nrel.gov

**ABSTRACT**

Growing energy storage demands on lithium-ion batteries (LIB) necessitate exploration of new electrochemical materials as next-generation battery electrode materials. In this work, we investigate the previously unexplored electrochemical properties of earth-abundant and tunable $Zn_{1-x}Sn_{1+x}N_2$ (x = -0.4 to x = 0.4) thin films which show high electrical conductivity and high gravimetric capacity for Li insertion. Although the goal of this work is to demonstrate electrochemical properties rather than cycle stability, enhanced cycling performance is achieved compared to previously published end-members $Zn_3N_2$ and $Sn_3N_4$, showing decreased irreversible loss and increased total capacity and cycle stability. The average reversible capacity observed is >1,050 mAh/g for all compositions and 1,220 mAh/g for Zn-poor (x = 0.2) films. Extremely Zn-rich films (x = -0.4) show improved adhesion, however Zn-rich films undergo a phase transformation on the first cycle. Zn-poor and stoichiometric films do not exhibit significant phase transformations, which often plague nitride materials, and show no required overpotential at the 0.5 V plateau. Cation composition x is explored as a mechanism for tuning relevant mechanical and electrochemical properties such as capacity, overpotential, phase transformation, electrical conductivity, and adhesion. The lithiation/delithiation experiments confirm the reversible electrochemical reactions. Without any binding additives (due to the exploratory nature of this work) the as-deposited electrodes delaminate resulting in fast capacity degradation. We demonstrate the mechanical nature of this degradation through decreased electrode thinning, resulting in cells with improved cycling stability due to increased mechanical stability. Combining composition and electrochemical analysis, this work for the first time demonstrates composition dependent electrochemical properties for the ternary $Zn_{1-x}Sn_{1+x}N_2$ and proposes earth-abundant ternary nitride anodes for increased reversible capacity and cycling stability.

**INTRODUCTION**

The rapid growth of global energy demand necessitates highly efficient and dense energy storage. Despite shortcomings in lithium-ion battery (LIB) technology, the expanding electric vehicle (EV) market along with growing demand for long-lasting portable electronics has established LIBs as the primary energy storage device. To compete with gasoline-powered vehicles, EV batteries need significantly increased energy-density and lower cost compared to state-of-the-art LIBs. Current LIBs use graphite anodes for superior cycle stability at the cost of energy density (372 mAh/g, 975 mAh/cm$^3$). Ideal and rigorously-researched anode solutions are Li-metal (3 860 mAh/g, 2 061 mAh/cm$^3$) and Si (3 580 mAh/g, 8 365 mAh/cm$^3$), however these technologies have failed to make an economic impact due to the formation of dangerous Li dendrites (Li-metal) and detrimental volume expansion of almost 300% (Si).[1,2] Uniform and pinhole-free coatings that are not reduced by Li are required to stabilize Li-metal against dendrite formation and unstable solid-electrolyte interfaces (SEI). An ideal coating is electrically insulating, ionically conducting, and robust against volumetric expansion/contraction, preventing reaction between Li and electrolyte while allowing Li-ions to pass through.[1] Another alternative is to use earth-abundant high-capacity alloy materials which are either natively stable against Li reduction or natively forms a passivating and ion-conducting SEI. This work examines the ternary $ZnSnN_2$ as such a potential next-generation high-capacity material.

Nitrides are, in general, of interest to the LIB community because of the Li-ion conducting properties of the base compound $Li_3N$ and often non-toxic nitride chemistry. The primary phase α-$Li_3N$ (space group 191, P6/mmm) consists of edge-sharing hexagonal lithium in-plane centered on one nitrogen atom. Two additional lithium atoms are located above and below, creating a hexagonal bipyramid with 8-coordinate nitrogen.[3] $Li_3N$ is natively Li deficient by ~1-2%, enabling an ultrafast intraplanar Li-ion hopping mechanism with an experimentally measured Li-ion conductivity of $10^{-3}$ $\Omega^{-1}$-$cm^{-1}$ at room temperature[4], one of the highest Li-ion conductivity values measured in a room temperature single crystal. Related structures possibly exhibiting similar conductivity values and partial conversion of the host material to $Li_3N$ are reasons often cited to pursue nitride negative electrodes. Nitrides generally show high gravimetric capacity, high thermal and chemical stability, and a low Li-insertion potential.[5–7]

There are two general categories of nitrides studied as negative electrodes, lithiated transition-metal nitrides and single metal nitrides. Examples of studied lithiated transition-metal nitrides include $LiMoN_2$[8], $Li_7MnN_4$,[9] and $Li_{3-x}M_xN$ where M is an element or combination of Co, Cu, Fe, and Ni[10–13]. Binary nitrides studied as negative electrodes include those containing Co,[6,14] Cr,[15,16] Fe,[14,16] Ge,[17] Mn,[18] Mo,[19] Ni,[20] Sb,[21] Sn,[22–24] Ti,[25] V,[26] and Zn[27,28]. Recent computational predictions suggest that binary nitrides and the corresponding lithium-containing ternaries are uniquely stable versus reduction by lithium when compared to oxides, sulfides, and fluorides of the same cation composition.[29] Separate experimental work has shown that Li-ternary nitrides have superior cycle stability at similar gravimetric capacity when compared to either Li-binary nitride components; i.e. $Li_{3-x-y}Co_xCu_yN$ is more stable than $Li_{3-x}Co_xN$ or $Li_{3-x}Cu_xN$, with an initial capacity of 500 mAh/g that increased to 590 mAh/g after 60 cycles.[30] A similar result is also demonstrated in the $Cr_xFe_{1-x}N$ system which showed an initial capacity 1800 mAh/g.[16] $SnN_x$ and $Zn_3N_2$ have both shown promising initial capacity values of 925 mAh/g and 1325 mAh/g, respectively, although neither report discusses cycle stability beyond initial capacity losses of 40% ($SnN_x$) and 55% ($Zn_3N_2$).[22,27] For $Zn_3N_2$, x-ray diffraction (XRD) experiments confirm that $Li^+$ insertion occurs in multiple stages where the initial conversion reaction forms $Li_xZn_y$ alloys in a matrix of $Li_3N$, followed by the reversible transition between the $Li_xZn_y$-$Li_3N$ matrix and a LiZnN end-member along with free $Li^+$.[27] The same reaction is theorized to occur for $SnN_x$.[22] Beyond Zn and Sn containing nitrides, a few other nitrides have shown outstanding capacity and/or cycle stability. Crystalline VN was deposited in thin film form on to glass substrates by reactive DC sputtering and loaded to 0.12 mg/cm², showing an initial capacity of 1500 mAh/g and a fairly stable reversible capacity around 800 mAh/g after 60 cycles.[26] Crystalline CrN was deposited on stainless steel substrates by reactive RF sputtering and loaded to 0.14 mg/cm², showing an initial capacity of 1800 mAh/g and a reversible capacity around 1100 mAh/g after 30 cycles.[15] Crystalline CoN nanoflakes were deposited by reactive RF sputtering on Cu-foil substrates and loaded to 0.12 mg/cm², showing an initial capacity of 1080 mAh/g with an initial reversible capacity of 760 mAh/g that increased to 990 mAh/g after 80 cycles.[6] For a more comprehensive list and table, please see the review article by *Balogun et. al. (2015)*.[5]

$ZnSnN_2$, an earth-abundant and non-toxic nitride, belongs to a family of II-IV-N semiconductors related to the widely studied III-N materials by replacing two group III cations with one group II and one group IV, resulting in an orthorhombic supercell with ordered cations (space group *Pna2₁*). II-IV-N compounds deposited at moderate temperatures typically form cation-disordered wurtzite (space group *P6₃mc*) structures, while low-temperature deposited materials are often amorphous. $ZnSnN_2$ in particular has been the subject of recent optoelectronics research due to its optical bandgap (1 eV – 2 eV), potential tunability, defect tolerance, earth-abundance, and non-toxic nature.[31,32] Cation composition x in $Zn_{1-x}Sn_{1+x}N_2$ has been previously shown to control conductivity, carrier density, carrier mobility, and optical absorption onset.[31] Cation composition in $Zn_{1-x}Sn_{1+x}N_2$ has been experimentally varied by combinatorial sputtering from x = -0.5 to x = 0.4,[31] indicating a large tolerance for oxidation state shifts and therefore a hypothetical tolerance for substantial lithium insertion. We theorize that $Zn_{1-x}Sn_{1+x}N_2$ will also exhibit electrochemical tunability as LIB electrodes considering the previously discussed $SnN_x$ and $Zn_3N_2$ results.[22,27]

In the context of other successful ternary and higher nitride alloys this work explores earth-abundant and non-toxic $Zn_{1-x}Sn_{1+x}N_2$ as a negative electrode for lithium-ion insertion, offering a reversible capacity of >1,200 mAh/g. Cation composition x in $Zn_{1-x}Sn_{1+x}N_2$ is used to tune electronic conductivity, mechanical adhesion, and electrochemical properties. Zn-rich films exhibit superior stability, stoichiometric films exhibit minimal phase transformation and overpotentials, and Zn-poor (x = 0.2) films exhibit the highest electrical conductivity.

**METHODS**

$Zn_{1-x}Sn_{1+x}N_2$ electrodes were prepared on solvent-cleaned (acetone and isopropyl alcohol) 2-inch square copper foil current collectors (25 μm thickness) at room temperature by RF magnetron co-sputtering from elemental Zn and Sn targets in an argon atmosphere, using a separate microwave electron cyclotron resonance (ECR) plasma source for reactive nitrogen. Films were initially deposited at three cation compositions targeting Zn-poor, stoichiometric, and Zn-rich films with a target thickness of approximately 250 nm. The Zn target power was varied from 1.68 to 2.17 W/cm² and the Sn target power was varied from 3.80 to 4.54 W/cm² to control composition, while the ECR source was operated at 150 W forward power. All growths were performed with a rotating platen to ensure film uniformity. Additional samples were deposited at extremely Zn-poor and Zn-rich conditions, and additional near-stoichiometric films were deposited at thicknesses of 90 nm and 130 nm. Control samples were grown alongside the foil substrates on masked Corning Eagle XG boro-aluminosilicate glass substrates. Composition was measured by x-ray fluorescence (XRF) as Zn percent of the total cation composition, assuming no cation vacancies, averaging 16 points across the nominally homogeneous film for error correction. A previously determined correlation between XRF intensity and SEM cross-sectional thickness was used to determine film thickness and calculate total electrode volume. The masked control samples on glass substrates were measured by profilometer to corroborate the thickness measurements, also averaging 16 points across the control wafer.

Our group has, previously investigated, in significant detail, the crystallization and properties of wurtzite $Zn_{1-x}Sn_{1+x}N_2$.[31,33–35] In contrast with some of those previous experiments, we have intentionally targeted amorphous films by depositing at room temperature to remove periodic potential barriers that can impede ion conductivity. The films were structurally interrogated using a Rigaku SmartLab diffractometer equipped with a Cu K$\alpha$ X-ray source and a 4-circle goniometer. To measure electrical conductivity, four-point probe measurements were taken at 16 points across the nominally homogeneous control sample (glass substrate for electrical isolation), with all samples displaying ideal ohmic behavior. The conductivity measurements were in excellent agreement with previously published work.[31]

To determine the planar density of the films, Rutherford backscattering spectrometry (RBS) was performed using a 2 MeV He$^+$ beam energy using a model 3S-MR10 RBS system from National Electrostatics Corporation in a 168° backscattering geometry. The samples were measured until the total integrated charge delivered to the sample was 80 μC, and in cases where the same point was measured twice, the measurements were added together for a total of 160 μC. Compositions and planar densities were determined by fitting using the RUMP analysis software.[36] Substrates were measured separately from the films to accurately fit the background. All films analyzed by RBS contained 8 – 12 atomic percent oxygen, some of which is related to a surface oxide during storage and some of which is unintentionally incorporated during growth. Oxygen incorporation in $ZnSnN_2$ has been shown to suppress unintentional electron doping through benign neutral defect formation, while enhancing optoelectronic properties.[33,35,37]

**Table I.** Properties of as-deposited $Zn_{1-x}Sn_{1+x}N_2$ films. Error for thickness, composition, and conductivity is given as the standard deviation of 16 points across the 2 square inch wafer. Density error, for RBS measurements, is obtained from the RUMP model (see text).

| Targeted x | Thickness (nm) | $\frac{Zn\%}{Zn\% + Sn\%}$ | Measured x | Conductivity (S-cm$^{-1}$) | Measured Density (g-cm$^{-3}$) | Loading (mg/cm$^2$) |
|---|---|---|---|---|---|---|
| 0.4 | 111 ± 1 | 29.6 ± 0.89 | 0.41 | 7.1 ± 0.5 | 4.56 ± 0.21 | 0.051 ± 0.002 |
| 0.2 | 292 ± 6 | 37.6 ± 0.31 | 0.25 | 22.5 ± 2.2 | 5.80 ± 0.29 | 0.184 ± 0.011 |
| 0 | 323 ± 9 | 47.8 ± 0.29 | 0.04 | 12.9 ± 1.5 | 6.16 ± 0.35 | 0.199 ± 0.013 |
| -0.2 | 293 ± 9 | 59.2 ± 0.25 | -0.18 | 9.2 ± 1.3 | 6.32 ± 0.35 | 0.170 ± 0.010 |
| -0.4 | 96 ± 4 | 72.0 ± 0.14 | -0.44 | 1.1 ± 0.2 | 6.01 ± 0.43 | 0.058 ± 0.005 |

The coin cells were assembled with our proposed ternary electrode and the Li-metal counter electrode in an Ar-filled glovebox. Note that all ternary nitride electrodes were directly deposited on the 25 μm copper current collector without any polymer binder and carbon conductive additives. 1.2 M $LiPF_6$ in 3:7 ethyl carbonate (EC) : ethyl methyl carbonate (EMC) by weight was used as the electrolyte with a Celgard separator. All cells were allowed to rest for 10 hours while monitoring the half-cell potential prior to the electrochemical tests. Constant current was applied during lithiation and delithiation between the voltage range of 0.01 V to 1.5 V. Cycling performance were carried out using a Maccor battery test station. The cell assembling was performed in an Ar-filled glove box and tested at room temperature.

**RESULTS & DISCUSSION**

The composition, conductivity, thickness, and density of the as-deposited electrodes are reported in Table I. Based on this data, the x = 0.4 sample is likely an outlier due to deviations in density and conductivity. XRD measurements reveal primarily amorphous films with barely discernable polycrystalline wurtzite patterns on the same order as the noise floor (Figure S1).

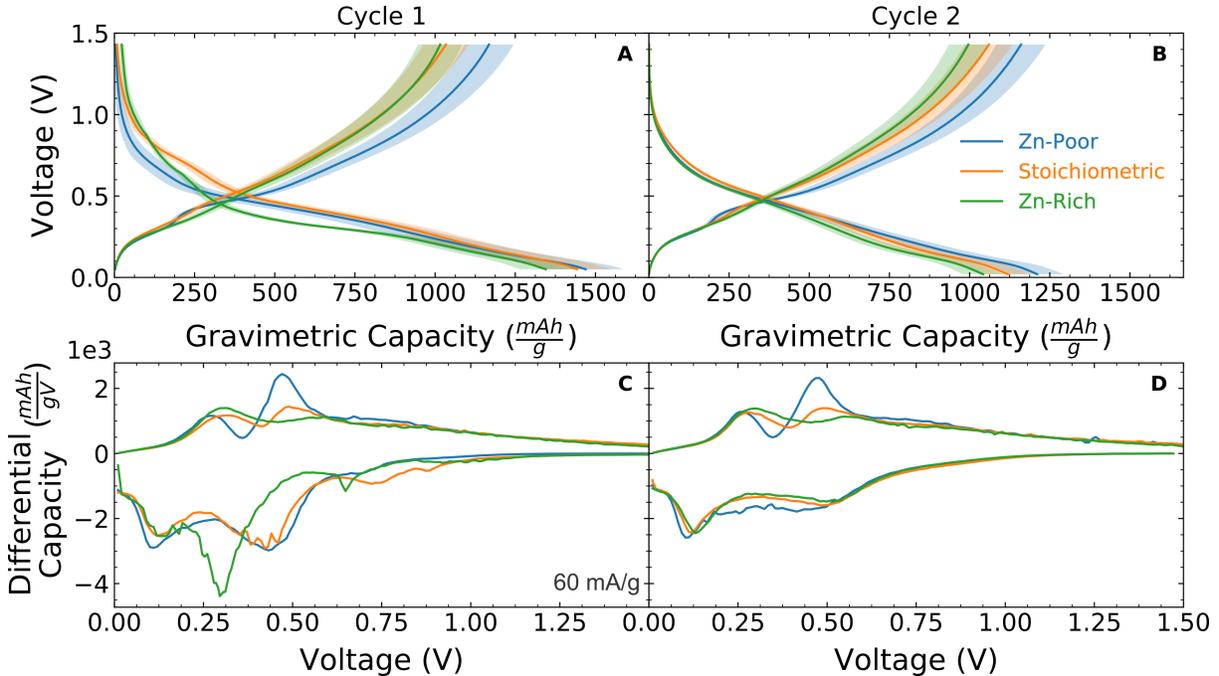

Figure 1. Voltage profiles (A-B) and dQ/dV plots (C-D) for the first two cycles of Zn-poor (x = 0.2), stoichiometric (x = 0), and Zn-rich (x = -0.2) electrodes cycled at 60 mA/g. (A): Average first cycle lithiation capacities exceed 1,300 mAh/g for all electrodes, with shaded regions showing the mean squared error. (B): The reversible capacities for all electrodes exceed 1,000 mAh/g with a Coulombic efficiency >97%. (C): the 1st cycle dQ/dV plots indicates an irreversible reaction associated with a phase transformation in Zn-rich electrode. (D): the 2nd cycle dQ/dV plots demonstrate the highly reversible behavior for all electrodes. The Zn-rich electrode resembles the similar electrochemical reactions as those of Zn-poor and stoichiometric electrodes. Panes B and D plots confirm that the Li storage capacity reduces with increasing Zn content.

Cells were cycled at a rate of 60 mA/g from 1.5 V to 0.01 V, the first and second capacity cycle is shown both as measured and differentiated in Figure 1. The shaded error region shown represents the mean squared error (MSE) of capacity averaged over multiple cells, taking in to account the standard deviation in measured thickness and density (Table I) which together determine the electrode weight. MSE is omitted in the differential capacity plot for clarity. The initial lithiation in Figure 1A consists of three sloped plateaus, more easily visualized as peaks in the differential capacity plot in Figure 1C, in ranges approximately 0.10 V - 0.15 V, 0.30 V – 0.45 V, and 0.65 V – 0.80 V.

According to previous literature for $Zn_3N_2$ a plateau is observed at ~ 0.12V corresponding to the formation of Li-Zn alloys and at 0.8 V corresponding to the formation of a solid electrolyte

**Table II.** Cycle statistic for $Zn_{1-x}Sn_{1+x}N_2$ electrodes versus lithium. Error due to material variability is propagated from property measurements (Table I) and averaged over multiple cells to determine capacity error. 50-cycle capacity loss is, in this case, a measure of thin film delamination as no binders were used to stabilize the electrode.

| x | Initial capacity (mAh/g) | Initial capacity (mAh/cm³) | Reversible capacity (mAh/g) | Reversible capacity (mAh/cm³) | Initial capacity loss (%) | 50 cycle capacity (mAh/g) | 50 cycle capacity (mAh/cm³) | 50 cycle capacity loss* (%) |
|---|---|---|---|---|---|---|---|---|
| 0.41 | 1791 ± 194 | 8172 ± 886 | 1192 ± 120 | 5439 ± 550 | 33 ± 9.9 | 657 ± 120 | 2996 ± 547 | 33 ± 16 |
| 0.25 | 1480 ± 52 | 9350 ± 325 | 1221 ± 38 | 7718 ± 242 | 17 ± 3.9 | 207 ± 56 | 1310 ± 358 | 80 ± 5.8 |
| -0.04 | 1497 ± 61 | 9228 ± 375 | 1131 ± 19 | 6974 ± 115 | 24 ± 3.3 | 599 ± 28 | 3693 ± 171 | 40 ± 2.9 |
| -0.18 | 1322 ± 134 | 8335 ± 848 | 1066 ± 12 | 6722 ± 74 | 19 ± 8.3 | 538 ± 173 | 3392 ± 1093 | 44 ± 18 |
| -0.44 | 1427 ± 93 | 6141 ± 402 | 1137 ± 64 | 4895 ± 277 | 20 ± 6.9 | 717 ± 52 | 3088 ± 226 | 14 ± 17 |

*50 cycle capacity loss calculated with respect to 7th cycle. Cycles 1-5: 60 mA/g, remaining cycles 120 mA/g

interface (SEI).[27] Detailed *in situ* XRD analysis in that work shows the formation of $LiZn_2$ and $Li_2Zn_3$ at 0.15 V and $LiZn$ at 0.05 V. For $Sn_3N_4$ and $SnN$ there are mixed and less detailed literature reports, with *Park et. al.*[23] reporting a reversible reaction around 0.2 V and an unnamed irreversible reaction around 0.9 V, while *Baggetto et. al.*[22] report that an irreversible conversion reaction determined not to be SEI formation occurs at 0.7 V and another quasi-plateau around 0.2 V – 0.3 V.

For $Zn_{1-x}Sn_{1+x}N_2$, the differential capacity peak (Figure 1C, during lithiation) around 0.75 V is irreversible, as is the large peak at 0.3 V visible primarily in the Zn-rich film. This peak accounts for most of the initial capacity loss in the Zn-rich sample. For the Zn-poor and stoichiometric cases the initial capacity loss primarily occurs during the reversible reaction around 0.45 V, with the same peak reversibly observable at smaller magnitude as shown in Figure 1D. The reduction peak around 0.1 V - 0.15 V is likely a summation of components from Li-Sn alloys and Li-Zn alloys, based on previous literature,[22,27] with a visible shift towards greater voltage with increasing Zn content. The same trend is observed for all lithiation and delithiation peaks, shifting to higher voltage for higher Zn content and broadening the peaks. We hypothesize that the voltage shifts are due to the composition change in the electrode material. Qualitatively the Zn-poor case contains features observed in $Sn_3N_4$ and the Zn-rich case shows features observed in $Zn_3N_2$, indicating that $Li_xSn_yN_z$, LiZnN, and $Li_3N$ are likely reversible end members for the conversion reaction as seen in previous literature.[22,27]

During delithiation, the reversible oxidation trend shows two clear peaks, one at 0.25 V – 0.3 V, and a second at 0.45 V – 0.55 V. The separation between these peaks/plateaus in the delithiation voltage profile for the Zn-poor film clearly resembles the voltage profile observed previously in $Sn_3N_4$ and SnN films, indicating that Zn-poor $ZnSnN_2$ tends towards tin nitride-like behavior as expected. Finally, the higher voltage redox couple around 0.45 V – 0.55 V shows nearly zero overpotential on oxidation, indicating that in this cell geometry the reaction is occurring at maximum efficiency with no resistive losses.

While the individual reactions of Li with Zn and Li with Sn are difficult to deconvolve, the likely mechanisms are proposed below based on previous research.[5,22–27] The initial reduction involves the direct conversion of $ZnSnN_2$ to Li-metal alloys in a matrix of $Li_3N$. The discharge plateau at 0.5 V is attributed to the formation of $Li_3N$, while the lower voltage discharge plateaus are attributed to the formation of the Li-metal alloys which trend towards higher voltages for higher Zn-compositions. Delithiation occurs by converting the Li-metal alloys to Zn and Sn around 0.3 V and the conversion of the metals with $Li_3N$ into Li-metal-N end members around 0.5 V, which represents the reversible reaction.

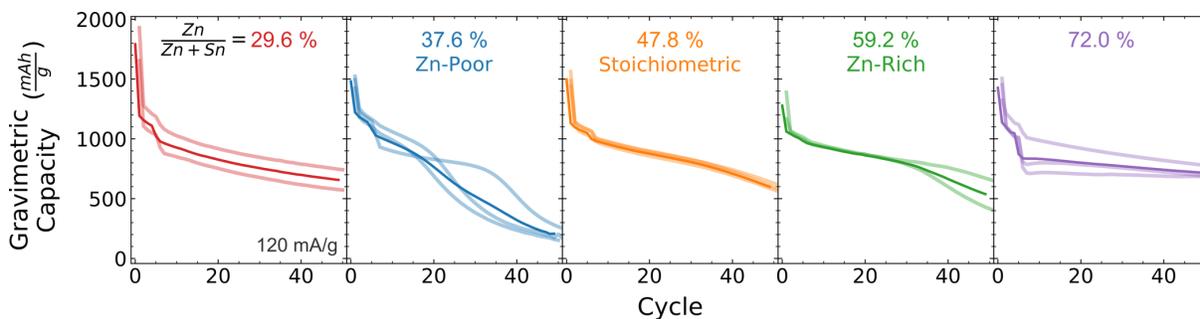

Figure 2. Mechanical stability (delamination) of the binder-free electrodes as a function of composition: Zn-poor (blue), stoichiometric (orange), and Zn-rich (green) $Zn_{1-x}Sn_{1+x}N_2$, along with extremely Zn-poor (red) and Zn-rich (purple) films. Pale lines are individual cell data, while darker lines represent the mean. The primary capacity loss mechanism is volume loss due to delamination, corroborated by post-mortem SEM. Zn-poor electrodes started to degrade at an enhanced rate after 20 – 30 cycles, while Zn-rich electrodes show the same mechanism after 30 – 40 cycles. The first 5 cycles were conducted at 60 mA/g with subsequent cycles at 120 mA/g. Extremely Zn-rich films with $\frac{Zn}{Zn+Sn}$ = 72% showed significantly improved cycle statistics, likely due to improved adhesion. Extremely Zn-rich and Zn-poor films here are thinner than the others (see **Table I**) which is partially responsible for increased stability.

Multiple cells were cycled at 120 mA/g for 50 cycles to observe cycle loss and mechanical stability, shown in Figure 2, and the average statistics are shown in Table II. For Figure 2, individual cells are shown as pale lines while the darker line represents the mean. In this plot we include films with extreme off-stoichiometry, x ~ ± 0.4, to demonstrate the wide range of tunability. Extremely Zn-poor, x ≈ 0.4 (Zn-rich, x ≈ -0.4), films were observed to have similar differential capacity trends to the Zn-poor, x ≈ 0.2 (Zn-rich, x ≈ -0.2), films, respectively (Figure S2). Of note are the initial capacity losses, which are much lower than that of either endpoint $Zn_3N_2$ (55%) or $Sn_3N_4$ (40%) with improved reversible gravimetric capacity.[22,23,27] It is notable that the extremely Zn-rich and Zn-poor films (x ≈ ± 0.4) are thinner than the rest of the sample set, likely aiding in mechanical stability and improving the cycle statistics. As will be described below, the primary capacity loss mechanism is volume loss due to mechanical instability such as delamination. For extremely Zn-rich films with x ≈ -0.4 we observe increased cycle stability which is most likely indicative of increased adhesion to the current collector. Zn-poor films (x ≈ 0.2) exhibit a fast capacity decay, which is caused by the poor mechanical stability and electrode delamination. Regardless, cation composition is shown as a useful tool to tune thin-film mechanical stability. Note that all of electrodes were directly deposited on the current collector without any binding and conductive additives. The mechanical integrity and the intrinsic conductivity of the electrode materials dominate the electrochemical cycling stability. Despite the fast capacity degradation, our work focuses on the effects of composition on the reversible capacity of these ternary nitrides and evaluate the potential of using these earth abundant materials for LIB anodes.

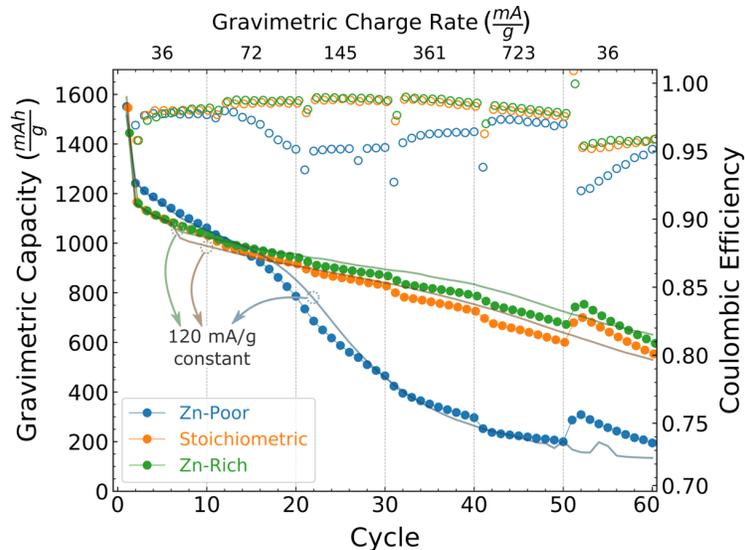

Figure 3. Rate performance from 36 mA/g to 723 mA/g for Zn-poor (blue), stoichiometric (orange), and Zn-rich (green) electrodes. Kinetically limited capacity steps are observed but minimal, likely due to the high electrical conductivity while implying sufficient ion conductivity. The capacity decay is attributed to materials loss caused by mechanical failures, as indicated in Figure 2. The cycling performance for all electrodes with a constant cycling rate of 120 mA/g is also plotted here (solid lines), overlaying the rate performance of the electrodes to demonstrate the root cause for the capacity decay. The closed circles are gravimetric capacity, while the open circles are Coulombic efficiency.

Charge-rate dependent tests were also performed on the $Zn_{1-x}Sn_{1+x}N_2$ half-cells to evaluate the rate performance of these electrodes. Cells were cycled for 10 cycles each at 36 mA/g, 72 mA/g, 145 mA/g, 361 mA/g, and 723 mA/g, returning to the original rate of 36 mA/g for 10 final cycles, the results of which are shown in Figure 3. The trend is overwhelmingly dominated by the mechanical failures, especially for the Zn-poor electrodes. As mentioned above, the Zn-poor electrodes have poor mechanical adhesion and integrity, which causes the drastic capacity decay. We expect that the significantly improved cycling performance would be achieved with enhanced mechanical strength. Minimal capacity loss is observed at each current increase, indicating that kinetically limited charge transport is insignificant up to the maximum rate of 723 mA/g, approximately C/2 assuming the first cycle capacity represents the full capacity. These high rates are possible due to the high electrical conductivity observed in $Zn_{1-x}Sn_{1+x}N_2$. The small steps at each current increase in Figure 3 are an indication of fast carrier transport, both electrons and ions, and indicate favorable ion conductivity when considering the high electronic conductivity of $Zn_{1-x}Sn_{1+x}N_2$ (see Table I). It is likely that much higher charge rates can be achieved using $Zn_{1-x}Sn_{1+x}N_2$.

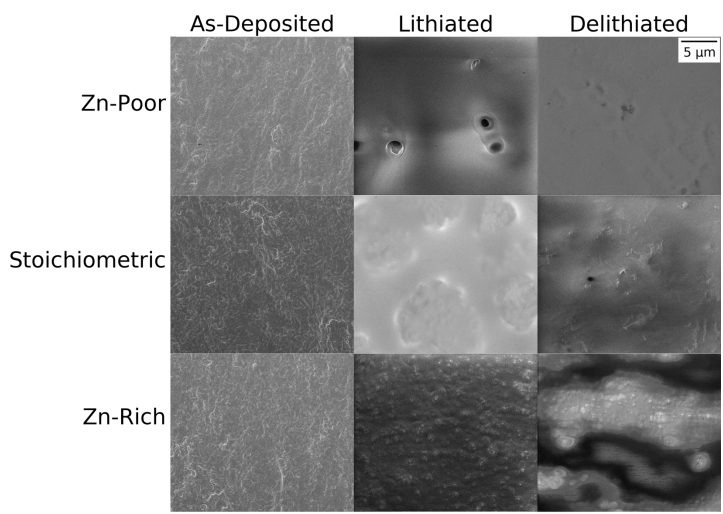

Figure 4. Top-view SEM micrographs of the as-deposited, lithiated and delithiated Zn-poor, stoichiometric, and Zn-rich $Zn_{1-x}Sn_{1+x}N_2$ electrodes. The lithiation and delithiation rate was set at 60 mA/g. Native films are porous, expanding to close the pores after lithiation due to volumetric expansion. Surface aggregation and roughening are observed upon lithiation, with the formation of pits in the Zn-poor case. Cycled films retain reduced porosity compared to the native films, with increased surface features. Cracks were not observed after cycling.

Scanning electron microscopy (SEM) was performed on the electrodes as-deposited, after one lithiation, and after one full cycle (delithiated) to observe changes in microstructure and monitor the electrodes for signs of cracking due to volumetric expansion. 50-cycle samples (post-mortem) were mostly devoid of analyzable material due to delamination. Samples were loaded into the SEM chamber immediately after coin cell disassembly to minimize exposure to atmosphere and observed at magnifications ranging from 100x to 100,000x. Cracks were not observed at any magnification level (Figure S3). The micrographs at 10,000x magnification are shown in Figure 4. The as-deposited films all appear similar, showing a porous microstructure with nanocrystalline grains (Figure 4 left column, also see Figure S2). Upon lithiation the porous structure disappears due to the volumetric expansion associated with insertion of 8 or more Li per formula unit. Pits are observed after lithiation in both Zn-poor and Zn-rich cases, with a much higher density of pits in the Zn-poor electrode. The lithiated stoichiometric and Zn-rich samples also show the formation of conglomerates on the surface, on the order of 5 μm for the stoichiometric film and 0.5 μm for the Zn-rich film. These conglomerates were not observed for the Zn-poor film. After delithiation the pore density in the Zn-poor and Zn-rich films is reduced, more so in the Zn-poor film, while pores appear in the stoichiometric film. From these images it appears that the primary mechanism for hosting the volumetric expansion and contraction is the formation and dissolution of conglomerates on the surface, although further studies on the chemical makeup of these features are required to confirm this hypothesis.

To further demonstrate the role of mechanical stability in capacity loss, near-stoichiometric samples were fabricated at thicknesses of 90 nm and 130 nm with loading masses of $0.047 \pm 0.003$ mA/cm$^2$ and $0.083 \pm 0.006$ mA/cm$^2$. The actual cation composition of these films (for stoichiometric films Zn/[Zn+Sn] = 50%) was between the Zn-poor and stoichiometric case as shown in Table I, with conductivity values similar to the Zn-poor film. The 50-cycle capacity trends of these films, compared with the thicker Zn-poor and stoichiometric films, are shown in

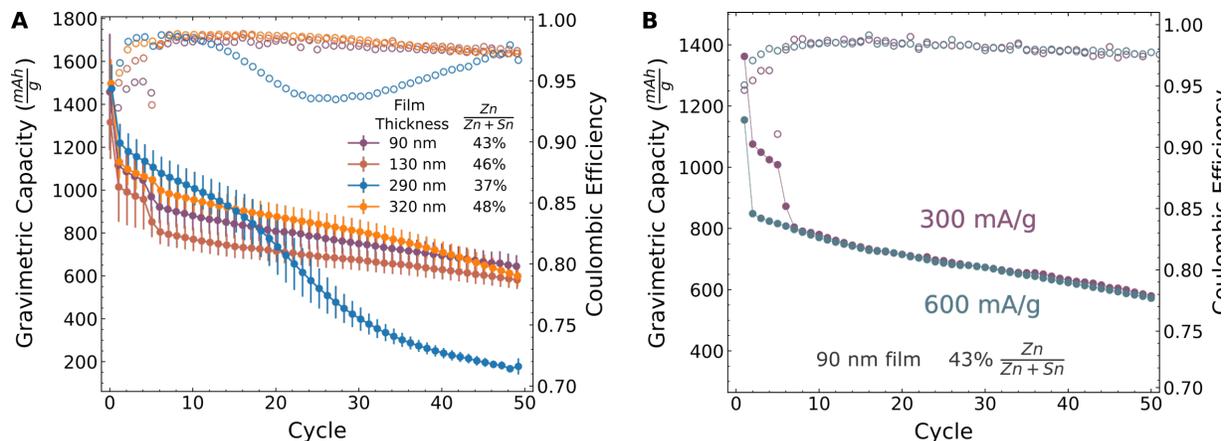

Figure 5. (A) Gravimetric capacity and Coulombic efficiency versus cycle for films of various thickness. Vertical bars indicate mean squared error across multiple cells and the electrode weight standard deviation. The first 5 cycles were cycled at 60 mA/g, while the remaining cycles occurred at 120 mA/g. Decreasing thickness is effective at improving mechanical stability and therefore cycle capacity. (B) lithiation/delithiation rate comparison for the 90 nm thick film showing no meaningful deviation between a cell cycled at 300 mA/g (first 5 cycles at 60 mA/g) and a cell cycled at 600 mA/g.

Figure 5A with error bars representing the MSE and weight uncertainty over multiple cells. Decreasing the film thickness is effective at significantly improving the 50-cycle stability loss. For thick stoichiometric (47.8% Zn/[Zn+Sn]) and Zn-poor (37% Zn/[Zn+Sn]) films the 50-cycle loss (compared to the 7th cycle where the rate is increased and constant) is 40% and 83%, respectively, while the 90 nm and 130 nm films show 28% and 30% loss, respectively. The larger loss in the 90 nm sample is likely due to the lower stoichiometry. Qualitatively, we observe a secondary slope onset after multiple cycles in thicker films which is not observed in the thinner films. Even after 100 cycles (Figure S4), this secondary slope onset is not observed for films > 200 nm. It is likely that even thinner films, less than 50 nm, will show significantly improved cycle statistics. Comparing this result to the cycle stability of the extremely Zn-rich and Zn-poor ($x \approx \pm 0.4$) films in Figure 2, we note that the extremely Zn-rich ($x \approx -0.4$) film exhibits superior mechanical stability compared to the $x \approx 0$, $x \approx 0.2$, and $x \approx 0.4$ films. Taking both the thickness and composition trends in context, we can conclude that increasing the mechanical stability through thinning results in improved cycle performance, and we can also independently conclude that modifying the composition towards a Zn-rich stoichiometry (x < 0) leads to improved mechanical stability as observed by cycle performance.

To demonstrate the rate capability of this material we also cycled the 90 nm film at a rate of 600 mA/g without any initial slow-rate equilibrium step. This data, compared with slower cycled data on films from the same deposition, is shown in Figure 5B. Here we observe no change in cycle statistics other than decreased kinetically limited transport for the first 5 cycles when operating at 60 mA/g, further demonstrating the high-rate capability of this electrode.

**CONCLUSIONS**

In conclusion, the electrochemical properties of earth-abundant $Zn_{1-x}Sn_{1+x}N_2$ have been explored in the context of LIBs. $Zn_{1-x}Sn_{1+x}N_2$ was deposited on Cu foil substrates with x ranging from 0.4 to -0.4. The films show high electrical conductivity of 1 S-cm$^{-1}$ – 20 S-cm$^{-1}$, appropriate for low resistive losses in electrochemical applications. When fabricated into half-cells versus a lithium metal counter electrode, $Zn_{1-x}Sn_{1+x}N_2$ films show extraordinarily high reversible gravimetric capacities exceeding 1,000 mAh/g across two redox couples. The differential capacity plots indicate that the ternary nitrites experience stepwise lithiation/delithiation with the lithiation products of $Li_xSn_yN_z$, LiZnN, and $Li_3N$. The first redox couple, at 0.5 V, shows minimal overpotential indicating a highly efficient conversion reaction with negligible resistive losses. Zn-rich films show a large phase transformation upon initial cycling, while stoichiometric and Zn-poor films do not. Along with other lithiation products, the formation of $Li_3N$ can enhance the ion diffusion and surface stability, leading to their excellent rate preperformance. The electrodes in our work were directly deposited on the current collector without any binder additives and the volumetric changes caused by lithiation/delithiation result in materials loss and delamination, and we attribute significant capacity loss to this mechanical instability. While the goal of this work is to explore and investigate the electrochemical properties of these earth abundant ternary nitrides, we have also investigated the cycle behavior as a proxy for volumetric loss. We have also fabricated and characterized thinner electrodes with increased mechanical stability, resulting in a maximum average 50-cycle capacity of 720 mAh/g for extremely Zn-rich ($x \approx -0.4$) material. Increasing the Zn fraction is also shown to be effective at preventing delamination, with a 50-cycle capacity loss of 14%. Overall, this work demonstrates for the first time the reversible electrochemical behavior of the ternary nitride $Zn_{1-x}Sn_{1+x}N_2$ and demonstrates the effects of high electrical conductivity and the formation $Li_3N$ on rate performance. The present results provide motivation to explore new earth-abundant nitrides as LIB anode materials.

SUPPLEMENTARY INFORMATION

Additional experimental details and characterization available as Figures S1-S4.

CONFLICTS OF INTEREST

Authors *Tellekamp, Osella, Tamboli, and Ban* have patent *US 2021/0036322 A1 published*.

ACKNOWLEDGMENT

C.Ban would like to thank financial support from the Research & Innovation Seed Grant Program at University of Colorado Boulder. M. B. Tellekamp would like to thank Dr. Yun Xu for helpful discussions, and Bobby To for acquiring SEM images. This work was authored in part by the National Renewable Energy Laboratory, operated by Alliance for Sustainable Energy, LLC, for the U.S. Department of Energy (DOE) under Contract No. DE-AC36-08GO28308. Funding provided by U.S. Department of Energy, Office of Science, Basic Energy Sciences. The views expressed in the article do not necessarily represent the views of the DOE or the U.S. Government.

DATA AVAILABILITY STATEMENT

The data that support the findings of this study are available from the corresponding author upon reasonable request.

REFERENCES

[1] D. Lin, Y. Liu, and Y. Cui, Nature Nanotechnology **12**, 194 (2017).
[2] M.N. Obrovac and L.J. Krause, J. Electrochem. Soc. **154**, A103 (2007).
[3] D.H. Gregory, The Chemical Record **8**, 229 (2008).
[4] U. v. Alpen, A. Rabenau, and G.H. Talat, Appl. Phys. Lett. **30**, 621 (1977).
[5] M.-S. Balogun, W. Qiu, W. Wang, P. Fang, X. Lu, and Y. Tong, Journal of Materials Chemistry A **3**, 1364 (2015).
[6] B. Das, M.V. Reddy, P. Malar, T. Osipowicz, G.V. Subba Rao, and B.V.R. Chowdari, Solid State Ionics **180**, 1061 (2009).
[7] N. Pereira, L. Dupont, J.M. Tarascon, L.C. Klein, and G.G. Amatucci, J. Electrochem. Soc. **150**, A1273 (2003).
[8] S.H. Elder, L.H. Doerrer, F.J. DiSalvo, J.B. Parise, D. Guyomard, and J.M. Tarascon, Chemistry of Materials **4**, 928 (1992).
[9] E. Panabière, N. Emery, S. Bach, J.-P. Pereira-Ramos, and P. Willmann, Electrochimica Acta **97**, 393 (2013).
[10] J.L.C. Rowsell, V. Pralong, and L.F. Nazar, J. Am. Chem. Soc. **123**, 8598 (2001).
[11] Y. Liu, T. Matsumura, N. Imanishi, T. Ichikawa, A. Hirano, and Y. Takeda, Electrochemistry Communications **6**, 632 (2004).
[12] J. Yang, K. Wang, and J. Xie, J. Electrochem. Soc. **150**, A140 (2003).
[13] J. Yang, Y. Takeda, N. Imanishi, and O. Yamamoto, Electrochimica Acta **46**, 2659 (2001).
[14] Z.-W. Fu, Y. Wang, X.-L. Yue, S.-L. Zhao, and Q.-Z. Qin, J. Phys. Chem. B **108**, 2236 (2004).
[15] Q. Sun and Z.-W. Fu, Electrochem. Solid-State Lett. **10**, A189 (2007).
[16] Q. Sun and Z.-W. Fu, Electrochem. Solid-State Lett. **11**, A233 (2008).
[17] N. Pereira, M. Balasubramanian, L. Dupont, J. McBreen, L.C. Klein, and G.G. Amatucci, J. Electrochem. Soc. **150**, A1118 (2003).
[18] Q. Sun and Z.-W. Fu, Applied Surface Science **258**, 3197 (2012).
[19] D.K. Nandi, U.K. Sen, D. Choudhury, S. Mitra, and S.K. Sarkar, ACS Appl. Mater. Interfaces **6**, 6606 (2014).
[20] F. Gillot, J. Oró-Solé, and M. Rosa Palacín, Journal of Materials Chemistry **21**, 9997 (2011).
[21] Q. Sun, W.-J. Li, and Z.-W. Fu, Solid State Sciences **12**, 397 (2010).
[22] L. Baggetto, N.A.M. Verhaegh, R.A.H. Niessen, F. Roozeboom, J.-C. Jumas, and P.H.L. Notten, J. Electrochem. Soc. **157**, A340 (2010).
[23] K.S. Park, Y.J. Park, M.K. Kim, J.T. Son, H.G. Kim, and S.J. Kim, Journal of Power Sources **103**, 67 (2001).
[24] B.J. Neudecker and R.A. Zuhr, Proceedings of the Electrochemical Society **99**, 295 (2000).
[25] M.-S. Balogun, M. Yu, C. Li, T. Zhai, Y. Liu, X. Lu, and Y. Tong, Journal of Materials Chemistry A **2**, 10825 (2014).
[26] Q. Sun and Z.-W. Fu, Electrochimica Acta **54**, 403 (2008).
[27] N. Pereira, L.C. Klein, and G.G. Amatucci, J. Electrochem. Soc. **149**, A262 (2002).
[28] J.B. Bates, N.J. Dudney, B. Neudecker, A. Ueda, and C.D. Evans, Solid State Ionics **135**, 33 (2000).
[29] Y. Zhu, X. He, and Y. Mo, Advanced Science **4**, 1600517 (2017).

[30] K. Wang, J. Yang, J. Xie, and S. Zhang, Solid State Ionics 5 (2003).
[31] A.N. Fioretti, A. Zakutayev, H. Moutinho, C. Melamed, J.D. Perkins, A.G. Norman, M. Al-Jassim, E.S. Toberer, and A.C. Tamboli, J. Mater. Chem. C **3**, 11017 (2015).
[32] A.D. Martinez, A.N. Fioretti, E.S. Toberer, and A.C. Tamboli, J. Mater. Chem. A **5**, 11418 (2017).
[33] A.N. Fioretti, J. Pan, B.R. Ortiz, C. Melamed, P.C. Dippo, L.T. Schelhas, J.D. Perkins, D. Kuciauskas, S. Lany, A. Zakutayev, E.S. Toberer, and A.C. Tamboli, Mater. Horiz. **5**, 823 (2018).
[34] A.N. Fioretti, E.S. Toberer, A. Zakutayev, and A.C. Tamboli, 42nd IEEE Photovoltaic Specialist Conference (PVSC) 1 (2015).
[35] I.S. Khan, K.N. Heinselman, and A. Zakutayev, J. Phys. Energy **2**, 032007 (2020).
[36] N.P. Barradas, K. Arstila, G. Battistig, M. Bianconi, N. Dytlewski, C. Jeynes, E. Kótai, G. Lulli, M. Mayer, E. Rauhala, E. Szilágyi, and M. Thompson, Nuclear Instruments and Methods in Physics Research Section B: Beam Interactions with Materials and Atoms **266**, 1338 (2008).
[37] J. Pan, J. Cordell, G.J. Tucker, A.C. Tamboli, A. Zakutayev, and S. Lany, Advanced Materials **31**, 1807406 (2019).

*Supporting Information*

# Electrochemical Characterization of High-Capacity Earth Abundant ZnSnN$_2$ Negative Electrodes

*M. Brooks Tellekamp[1†], Anna Osella[1], Karen M. Heinselman[1], Adele C. Tamboli[1], Chunmei Ban[1,2]*

[1] National Renewable Energy Laboratory, 15301 Denver West Parkway, Golden, CO 80401

[2] Paul M Rady Department of Mechanical Engineering, University of Colorado Boulder, 1111 Engineering Dr, UCB 427, Boulder, CO, 80209

[†] brooks.tellekamp@nrel.gov

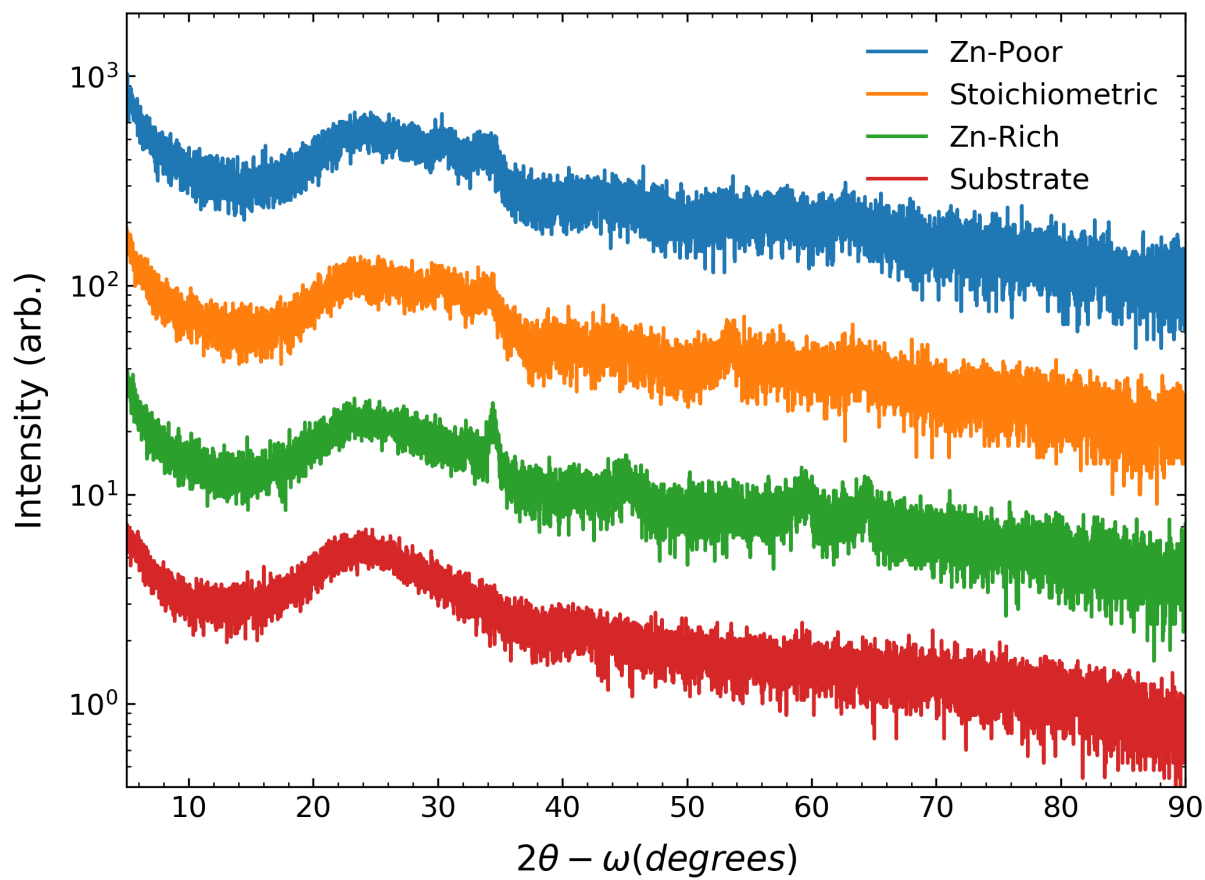

**Figure S1.** X-ray diffractogram of the as-deposited $Zn_{1-x}Sn_{1+x}N_2$ for x = 0.25 (Zn-poor), 0.04 (stoichiometric), and -0.18 (Zn-rich), along with the glass substrate.

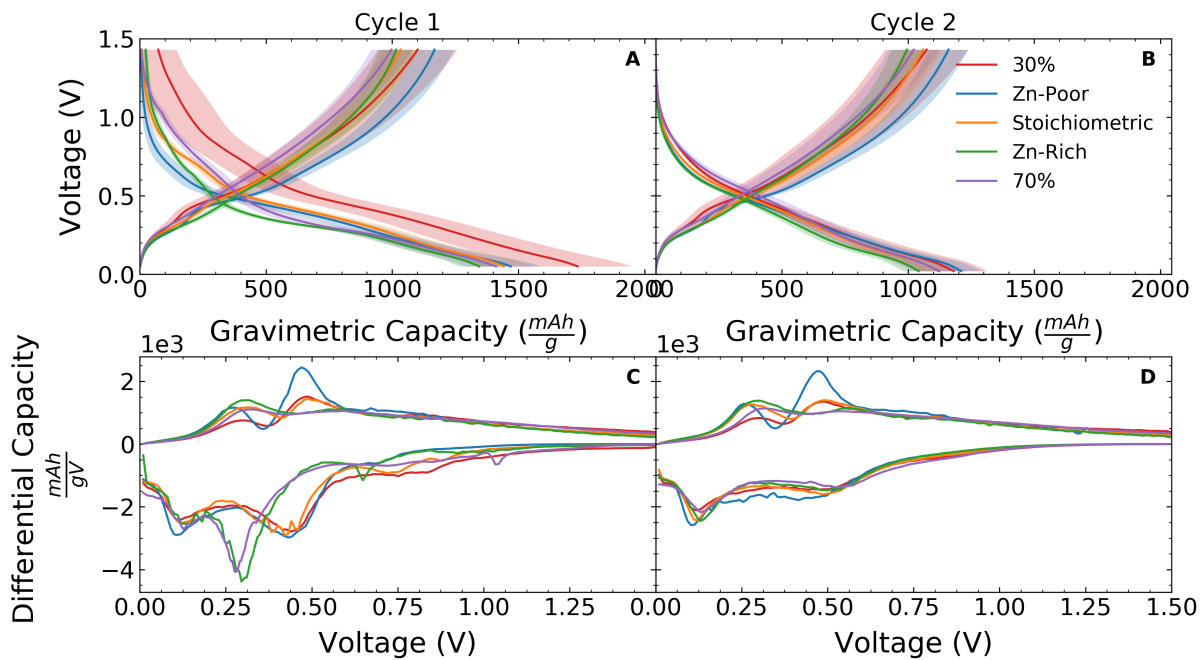

**Figure S2.** First (A, C) and second (B, D) cycle data for films with Zn/(Zn+Sn) = 30% (red) and 70% (purple), compared to the data in manuscript Figure 1. The extremely off-stoichiometric films share similar CV trends. The Zn-rich films (60% and 70%) both show a phase transformation and additional peak at 0.3V. The Zn-poor films (30% and 40%) also show similar CV curves.

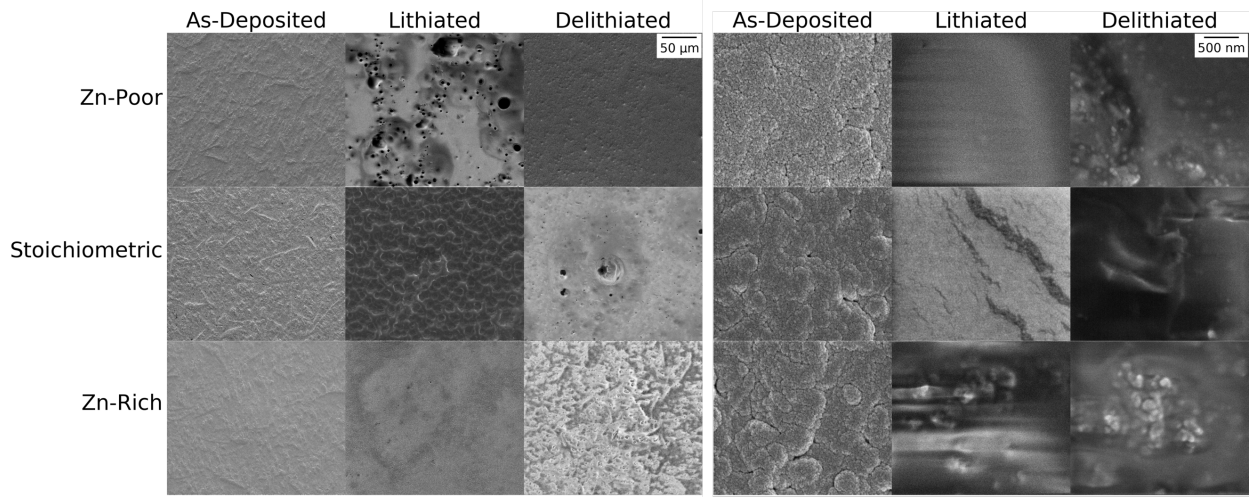

**Figure S3.** SEM images at 1,000x (left) and 100,000x (right) magnification.

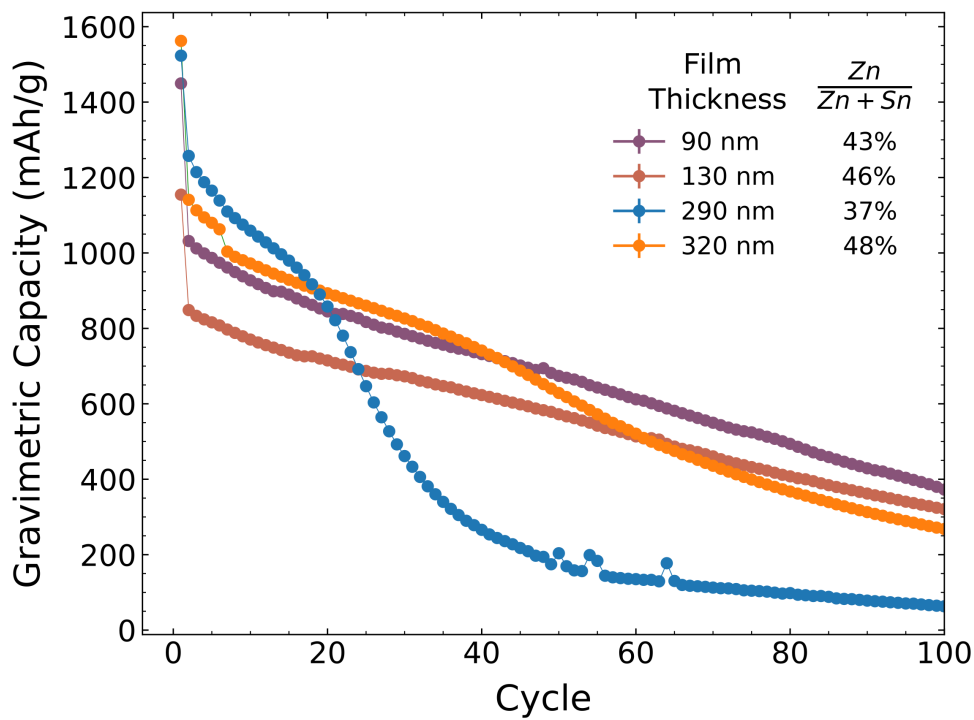

**Figure S4.** 100-cycle data for the film thickness series in manuscript Figure 5 demonstrating a constant slope for the thinner films, while the thicker films undergo a secondary loss.

24